\documentclass[11pt, a4paper]{article}

\usepackage{graphicx}
\graphicspath{{Figures/}}
\usepackage{layouts}
\usepackage{booktabs}
\usepackage{authblk}
\usepackage{amsmath}

\title{Transformation optics: a time- and frequency-domain analysis of electron-energy
loss spectroscopy}
\author[1]{Matthias Kraft \thanks{matthias.kraft09@imperial.ac.uk}}
\author[2]{Yu Luo \thanks{luoyu@ntu.edu.sg}}
\author[1]{J.B. Pendry}
\affil[1]{Blackett Laboratory, Department of Physics, Imperial College London, London SW7 2AZ, United Kingdom}
\affil[2]{School of Electrical and Electronic
Engineering, Nanyang Technological University, Nanyang Avenue 639798, Singapore}

\begin{document}
\maketitle
\begin{abstract}
Electron energy loss spectroscopy (EELS) and Cathodoluminescence (CL)
play a pivotal role in many of the cutting edge experiments in plasmonics.
EELS and CL experiments are usually supported by numerical simulations,
which, whilst accurate, may not provide as much physical insight as
analytical calculations do. Fully analytical solutions to EELS and
CL systems in plasmonics are rare and difficult to obtain. This paper
aims to narrow this gap by introducing a new method based on Transformation
optics that allows to calculate the quasi-static frequency and time-domain
response of plasmonic particles under electron beam excitation.
\end{abstract}

Electron energy loss spectroscopy (EELS) has always been at the heart
of plasmonics research, playing a major role in the experimental discovery
and characterization of plasmons \cite{Ritchie1957, Watanabe1956, Powell1959}.
Experimental and system design progress has been steep since then,
now allowing for an energy resolution of a few tens of meV, while
maintaining a sub-nanometer spatial resolution \cite{Krivanek2009}.
This makes EELS the ideal tool to study plasmons in metallic nano-particles.
Particularly, the high spatial resolution and ability to excite the
`dark' modes of a nano-particle provide advantages over optical methods.
This makes it possible to map out plasmon modes on a sub-nanometer
scale and gather information on the local density of states \cite{GarciaDeAbajo2008, Hohenester2009},
as has been shown in numerous experiments \cite{Nelayah2007, Nicoletti2011, Gomez-Medina2008, Chaturvedi2009, Koh2009, Koh2011}.
Quantum effects are negligible in plasmonic systems down to the nanometer
scale \cite{Duan2012}, however for separations between nano-particles
less than one nanometer quantum effects become important \cite{Scholl2013, Savage2012}.
Here too, EELS provides an effective method to probe this regime,
thanks to its high spatial resolution \cite{Scholl2013, Scholl2012, Duan2012}. 

Complementing EELS is the method of cathodoluminescence (CL). As in
EELS, fast electrons are used in CL to excite plasmons, however, instead
of measuring the energy loss experienced by the electrons, CL measures
the energy scattered by the nano-particle. This too, gives important
insights into the nature of plasmons and has found widespread application
in recent experiments \cite{Kuttge2009, Gomez-Medina2008, Coenen2012, Atre2015}.
A nice review on both EELS and CL can be found in \cite{GarciaDeAbajo2010}. 

On the theoretical side, these experiments are supported by a wide
range of numerical methods \cite{Duan2012, Cao2015, Matyssek2011, Talebi2013, Bigelow2012, GarciadeAbajo2002}.
Fully analytical methods are rare \cite{GarciaDeAbajo2010}, but have
been obtained in the non-retarded limit for geometries for which Poisson's
equation is separable as in, e.g. \cite{Illman1989, Zabala1997}.
This is where Transformation optics (TO) comes in, as it has been
shown to be a versatile method in the analytical study of plasmon
excitations in nano-particles of more complex shapes \cite{TOPlasmonics}.

TO is a relatively new tool for the study of Maxwell's equations that
has been developed over the last couple of decades \cite{WardTO, TOScienceReview12, TOReviewChen, TOReviewLeonhardt}.
Early research focused on the design of invisibility cloaks and other
functional devices such as beam splitters, beam shifters etc. \cite{PendryScienceCloak, LeonhardtCloak,  TOBeamSplitterRahm, TOPolarisationWerner, Huidobro2010, Liu2010}.
In recent years, TO has been applied as an analytical, rather than
a design tool and has been used to solve a range of problems in plasmonics
\cite{TOScienceReview12, TOScienceReview15, ZhangOEMultiFano}. Amongst
them, the interaction of closely spaced nano-particles \cite{AubryPRL},
non-locality in plasmonic nano-particles \cite{AntonioNonlocalTO}
and the calculation of van-der-Waals forces~\cite{PendryNatPhys, LuoPNAS2014}.
Most recently, TO has also been used to reveal hidden symmetries in
seemingly unsymmetrical structures \cite{Kraft2014}, calculate the
optical properties and dispersion relation of plasmonic metal gratings
\cite{Kraft2015}, as well as graphene gratings with periodically
modulated conductivities \cite{PalomaGraphene, PalomaMetasurface}.
More comprehensive reviews and introductions to TO as used in this
paper can be found in \cite{TOScienceReview12, TOScienceReview15, TOPlasmonics}.

Here we introduce a new application of TO; the analysis of electron
energy loss spectroscopy (EELS) problems for plasmonic nano-particles.
We will derive analytical expressions for the electron energy loss
and photon scattering spectra of a non-concentric annulus and an ellipse
(Supplementary Material), calculate the electrostatic potential in
frequency space and deduce the time-evolution of the system from this.

\section*{Transformation optics for EELS}

\begin{figure*}[t!]
\centering
\includegraphics[width=\textwidth]{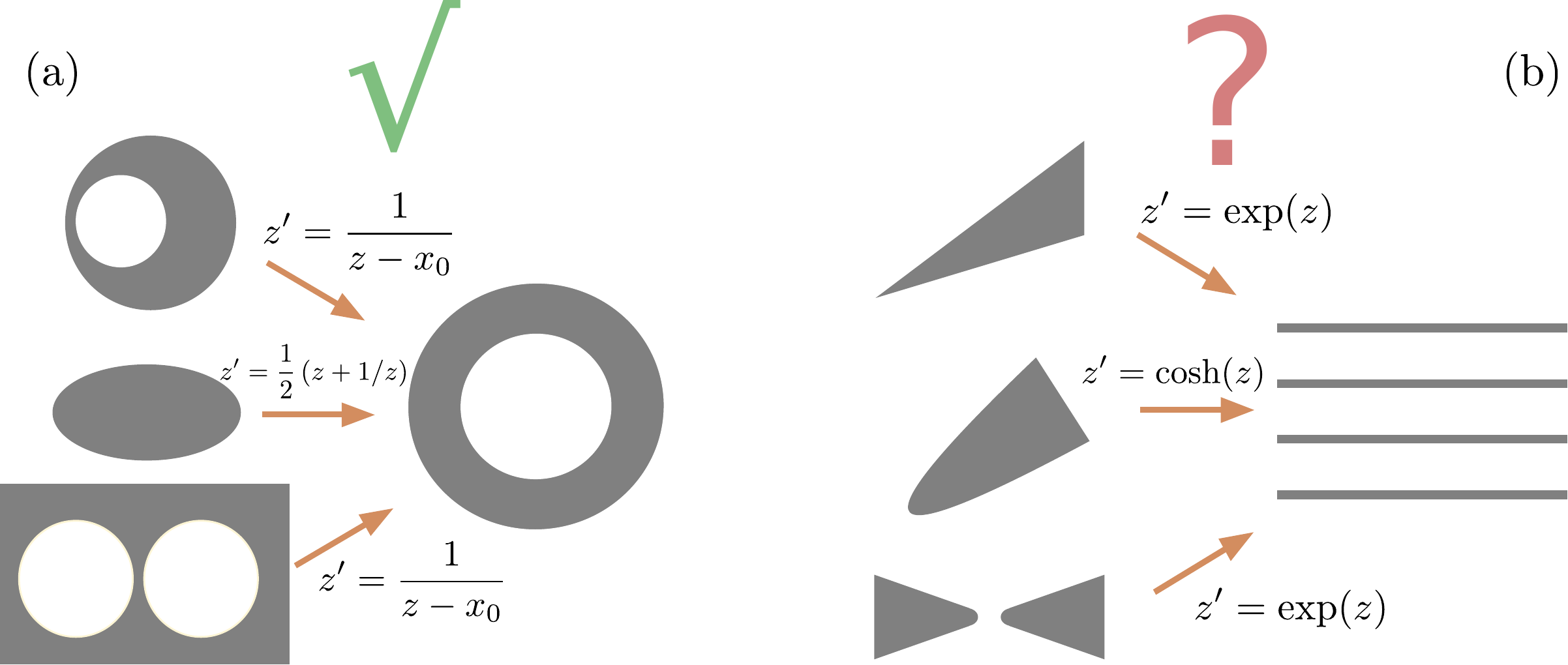}
\caption{Schematics of conformal transformations between related systems. Figure (a) shows a set of different geometries, for which a transformation optics analysis of EELS calculations is possible. The ellipse is studied in the supplemental material and results for the dimer are published elsewhere. Figure (b) shows some proposed structures for future applications of the TO approach to EELS calculations.} \label{fig:TO_schematics}
\end{figure*}

In a nutshell, TO works by relating two geometries via a coordinate
transformation. This is useful if a complicated geometry can be mapped
onto a simpler one with higher symmetry, as the higher symmetry often
simplifies analytical calculations and makes otherwise intractable
problems soluble. In the present case, we consider two-dimensional
problems and use conformal transformations \cite{Ablowitz} to analyze
our system. The great advantage of conformal transformations is that
they preserve the material properties of the structure, i.e. the in-plane
components of the permittivity and permeability tensor are invariant,
the electrostatic potential is invariant under these transformations,
too \cite{LuoYThesis}. Figure \ref{fig:TO_schematics} (a) and (b)
show examples of systems related via conformal maps (here $z=x+iy$
and $z'=x'+iy'$).

In figure \ref{fig:TO_schematics} (a) a non-concentric annulus, an
ellipse and two nearly touching cylinders can be related to a concentric
annulus. These structures have at most two mirror planes, but possess
a hidden rotational symmetry evident upon transformation to the concentric
annulus \cite{Kraft2014, TOPlasmonics}. The higher symmetry greatly
simplifies calculations of the plasmon eigenmodes of these systems
and, as it turns out, also allows us to derive analytical solutions
to these systems if an electron moves past. The non-concentric annulus
is analyzed in the main body of the paper, the ellipse is given in
the Supplementary Materials and results for the cylindrical dimers
are published elsewhere together with an analysis of two nearly touching
spheres \cite{YuEELS}. We would like to mention to the reader that
we have developed an open-source GUI software package implementing
the TO calculations for the three two-dimensional cases, calculating
electron energy loss and scattering spectra. It is available from
the authors.

Figure \ref{fig:TO_schematics} (b) shows another set of geometries,
a sharp knife-edge, a hyperbolic knife and a bow-tie antenna with
blunt tips, which can all be related to a system of periodic slabs
\cite{LuoNanoLett1, MiguelConfAnt}. These structures have also been
previously analyzed with TO, so we feel confident that EELS calculations
for these structures can be tackled with TO. We do not provide an
analysis of them here, but would like to highlight the generality
of the approach.
\begin{figure*}[b!]
\centering
\includegraphics[width=\textwidth]{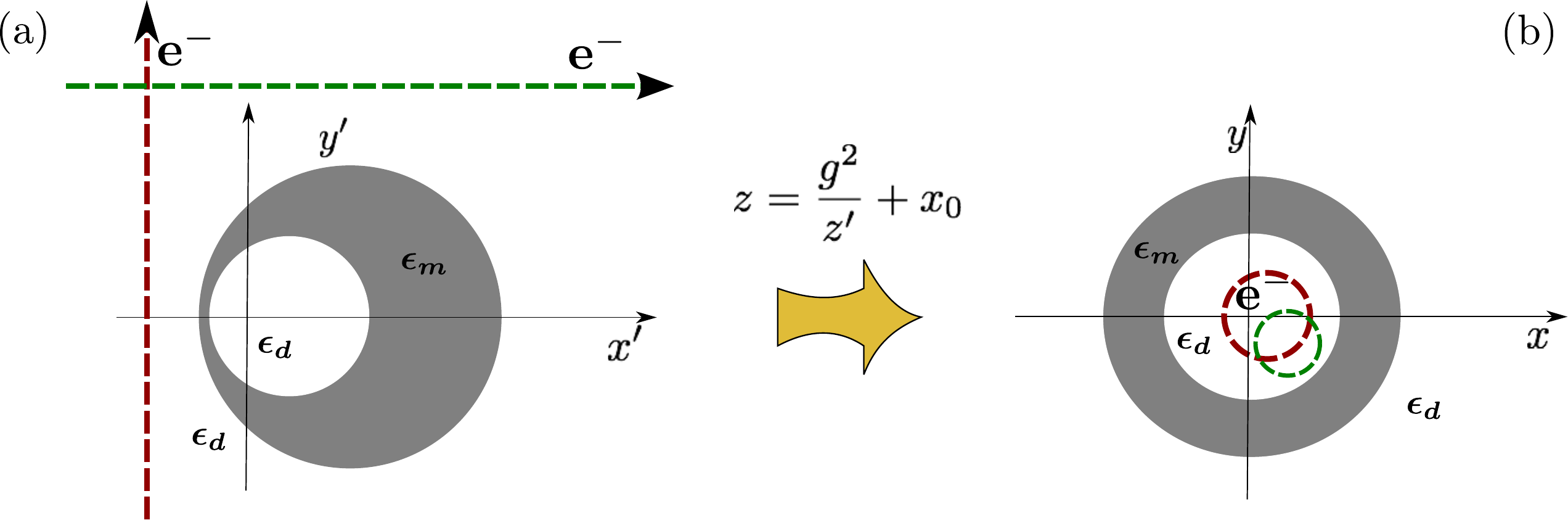}
\caption{Schematics of conformal transformations between related systems. Figure (a) shows the system under investigation in this paper. It can be related to the concentric annulus shown in (b). The simple geometry of the annulus allows for an analytical solution of the problem.} \label{fig:TO_annulus}
\end{figure*}

The remainder of this paper focuses on the non-concentric annulus.
A more detailed view of the problem considered here is shown in figure
\ref{fig:TO_annulus}. It shows the relation between the non-concentric
annulus frame (physical frame) and concentric annulus (virtual frame).
We consider a (line) electron moving past the non-concentric annulus
in the vertical (red line) and horizontal direction (green line).
Since charge is a conserved quantity, the transformation automatically
gives us the electron's trajectory in the virtual frame. There the
electron moves on a closed circular trajectory, however, it does so
with a non-uniform speed. This makes it difficult to calculate the
electrostatic potential of the electron in the virtual frame. In the
physical frame the electrostatic potential of the moving electron
is very simple to calculate, since the electron moves on a straight
line with constant velocity. We will thus adopt a hybrid approach
here. We will first calculate the source potential of the electron
in the original, physical frame, then transform it to the virtual
frame. We then calculate the response of the metal nano-particle in
the virtual frame of the concentric annulus and finally transform
the electrostatic potential back to the physical frame.

A word of caution is needed here. The calculation presented below
is strictly in 2-D, this means the moving electron really is a line
electron and all results are per unit length. In reality, the electron
is a point particle, so when it moves past the nano-particle it not
only loses momentum/energy in the in-plane direction, but also in
the out-of plane direction. This is not captured by the 2-D calculation,
however there is a workaround. The trick is to filter out the electrons
which have not lost any momentum out-of plane. Those effectively behave
as line charges. 

The electrostatic potential of an electron moving in a straight line
in the vertical direction at position $x_{e}'$ is easily deduced
as
\begin{align}
\phi^{sou}= & -\frac{\lambda}{4\pi\epsilon_{0}\omega}e^{-i\frac{\omega}{v}y'}e^{-|x'-x'_{e}|\frac{\omega}{v}}.\label{Eq:PotSou}
\end{align}
Here $c_{e}$ is the (line) electron's velocity, $\lambda$ is the
charge per unit length and both $\omega$ and $c_{e}$ are positive.
If the electron passes to the left of the non-concentric annulus,
the potential at the surface of the non-concentric annulus can be
written as
\begin{align}
\phi^{sou}= & \frac{\lambda\exp\left[\frac{\omega}{c_{e}}x_{e}\right]}{4\pi\epsilon_{0}\omega}\exp\left[-\frac{\omega}{c_{e}}(z')^{*}\right],
\end{align}
since $x_{e}'<x'$. The potential in the virtual frame is obtained
by substituting coordinates
\begin{align}
\phi^{sou}= & \frac{\lambda\exp\left[\frac{\omega}{c_{e}}x_{e}\right]}{4\pi\epsilon_{0}\omega}\exp\left[-\frac{\omega}{c_{e}}\left(\frac{g^{2}}{z-x_{0}}\right)^{*}\right]\\
= & \frac{\lambda\exp\left[\frac{\omega}{c_{e}}x_{e}\right]}{4\pi\epsilon_{0}\omega}\exp\left[-\frac{\omega}{c_{e}}\left(\frac{g^{2}}{re^{i\phi}-x_{0}}\right)^{*}\right],
\end{align}
which can be expanded into the eigenfunctions of the concentric annulus
\cite{YuEELS}
\begin{align}
\phi^{sou}= & \sum_{n=0}^{\infty}a_{n}^{s\pm}\left(\frac{r}{x_{0}}\right)^{\pm n}e^{\mp in\phi}.
\end{align}
Expressions for the expansion coefficients can be found in the Supplementary
Material. 

The form of the source potential suggests that the total electrostatic
potential in the virtual frame can be written as
\begin{align}
\phi_{I} & =\sum_{n=0}^{\infty}\left[a_{n}^{s+}e^{-in\phi}+(b_{n}+a_{n}^{rad})e^{in\phi}\right]\left(\frac{r}{x_{0}}\right)^{n}\quad & \mbox{for}\quad r<x_{0}\\
\phi_{II} & =\sum_{n=0}^{\infty}\left[(a_{n}^{s-}+a_{n}^{rad})\left(\frac{x_{0}}{r}\right)^{n}e^{in\phi}+b_{n}\left(\frac{r}{x_{0}}\right)^{n}e^{in\phi}\right] & \mbox{for}\quad R_{0}>r>x_{0}\\
\phi_{III} & =\sum_{n=0}^{\infty}\left[c_{n}\left(\frac{x_{0}}{r}\right)^{n}e^{in\phi}+d_{n}e^{in\phi}\left(\frac{r}{x_{0}}\right)^{n}\right] & \mbox{for}\quad R_{1}>r>R_{0}\\
\phi_{IV} & =\sum_{n=0}^{\infty}e_{n}e^{in\phi}\left(\frac{x_{0}}{r}\right)^{n} & \mbox{for}\quad R_{1}>r,
\end{align}
where $R_{0}$ and $R_{1}$ are the inner and outer radius of the
annulus, respectively. The coefficients $b_{n},c_{n},d_{n}$ and $e_{n}$
are the usual electrostatic scattering coefficients and can be determined
by demanding continuity of the tangential component of the electric
field and normal component of the displacement field at the interfaces.
The coefficients $a_{n}^{rad}$ are not present in a purely electrostatic
calculation; they encode information about the radiative reaction
of the system and have to be determined from an additional `radiative'
boundary condition \cite{Kraft2015, YuEELS}. Details can be found
in the supplementary. Using this additional boundary condition all
scattering coefficients can be uniquely determined.

\section*{Spectra for non-concentric annulus}

Energy is a conserved quantity. This means the energy absorbed by
the annulus in the virtual frame is the same as the energy absorbed
by the non-concentric annulus in the physical frame and similarly
for the power scattered by the nano-particles. 

The power absorbed by the annulus can be deduced from the resistive
losses in the metal, i.e. from
\begin{align}
Q= & \frac{1}{2}\int_{S}dS\mbox{Re}(\mathbf{j}^{*}\cdot\mathbf{E}),
\end{align}
with the integration area $S$ corresponding to the area of the annulus.
As the electrostatic potential is known, formulas for the electric
fields and currents are easily deduced and the integration readily
carried out (see Supplementary Material). After a straightforward
calculation the power absorbed by the annulus and hence also by the
non-concentric annulus is obtained as, 
\begin{align}
Q= & \pi\omega\epsilon_{0}\mbox{Im}(\epsilon_{m})\sum_{n=0}^{\infty}\left[n|c_{n}|^{2}x_{0}^{2n}(R_{0}^{-2n}-R_{1}^{-2n})+n|d_{n}|^{2}x_{0}^{-2n}(R_{1}^{2n}-R_{0}^{2n})\right].\label{Eq:PowAbs}
\end{align}
 The power scattered by the nano-particles can also be calculated
analytically. It can be obtained by imagining that the non-concentric
annulus is surrounded by a fictional absorber far away from the boundary
of the non-concentric annulus. One can then imagine that the power
scattered by the nano-particle is going to be equal to the power absorbed
by the fictional absorber. In the annulus frame this fictional absorber
corresponds to a point particle at position $x_{0}$ with non-zero
polarizability. The power scattered is then simply equal to the power
absorbed by the fictional point particle. Details on this approach
can be found in \cite{AlexRadLoss} and the Supplementary Material.
Carrying out the detailed calculation one arrives at the following
formula for the power scattered by the non-concentric annulus
\begin{align}
P_{sca}= & \frac{\pi^{2}\epsilon_{0}k_{0}^{2}g^{4}\omega}{2}\left|\sum_{n}\frac{nb_{n}}{x_{0}}\right|^{2}.\label{Eq:PowSca}
\end{align}

\begin{figure*}[t!]
\centering
\includegraphics[width=\textwidth]{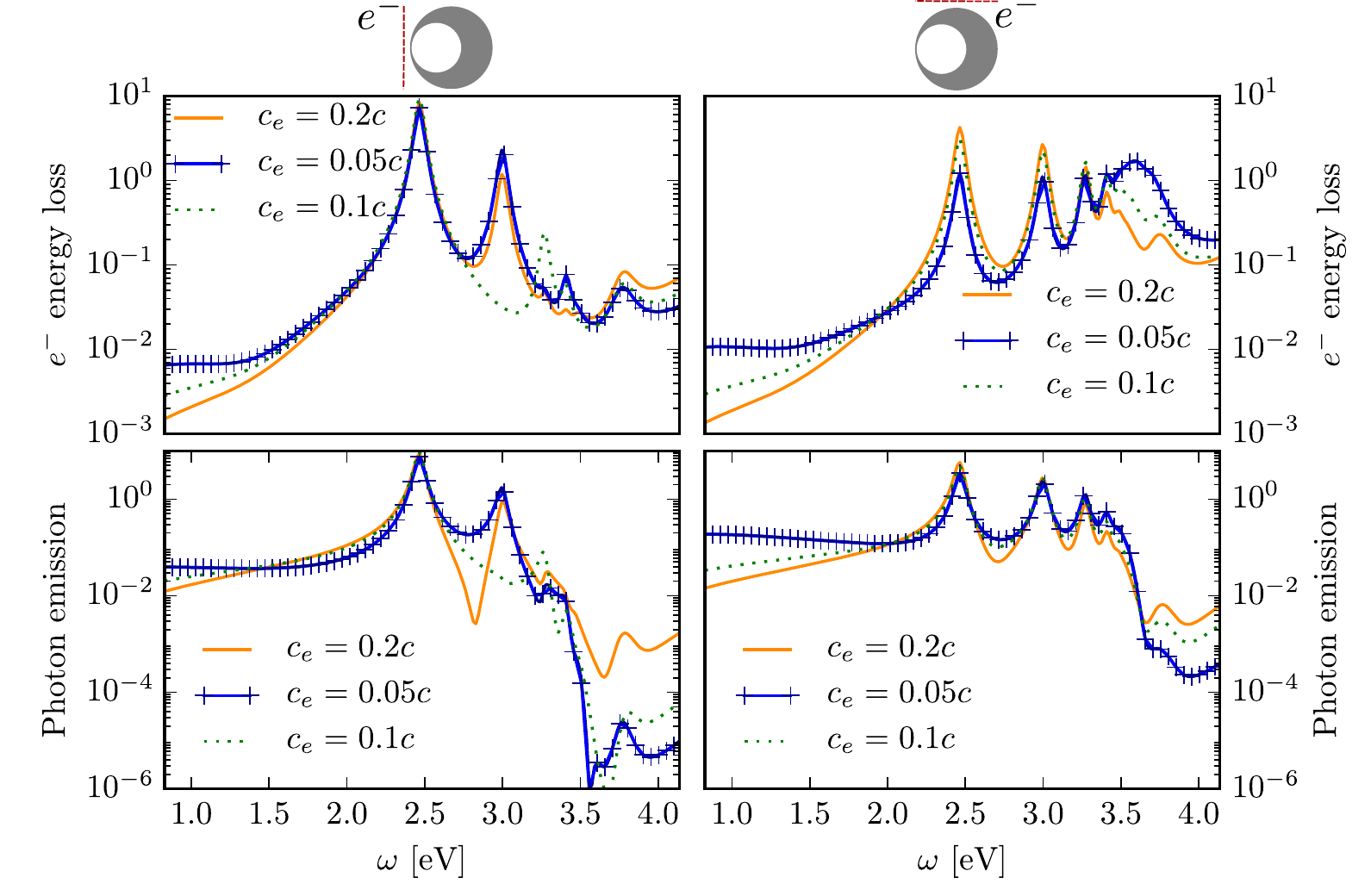}
\caption{Electron energy loss (top) and photon emission spectra (bottom) for an electron passing to the left (left) and electron passing along the top (right). The electron energy loss spectra give the energy loss probability density for the exciting electron. It is in units of [eV$^{-1}$ per electron per unit length]. The photon emission spectra give the photon emission probability density in units of [eV$^{-1}$ per electron per unit length]. Both curves are normalized such that the area under the curve is unity. The total energy lost by the electron and the total energy scattered over all frequencies is given in table \ref{tb:ELossLeft_cresc} for all the cases shown here. The geometrical parameters were set to $g^2=2\times10^{-8}$, $x_0=1.5$, $R_0=\exp(1)$ and $R_1=\exp(1.2)$. This gives a diameter of $\approx21nm$ for the outer cylinder and $\approx15nm$ for the diameter of the inner cylinder of the non-concentric annulus. The thickness at the thinnest point is $\approx0.6nm$. The distance of the electron to the particle was $\approx3.3nm$.}\label{fig:Spectra_subplot}
\end{figure*}

It is well known that moving electrons can probe both bright and dark
modes of plasmonic nano-particles (this is also one major advantage
over far field light excitation of plasmons). Eq.\ref{Eq:PowAbs}
and Eq.\ref{Eq:PowSca} are in accordance with this statement, as
they contain contributions from all plasmon modes of the system.

In EELS and cathodoluminescence experiments measurable quantities
are the total electron energy loss, i.e. the amount of energy an electron
passing the nano-particle loses and the photon scattering spectrum,
i.e. the amount of energy the nano-particle scatters. The total electron
energy loss is obtained by adding the power absorbed and power scattered
by the nano-particle. The photon scattering spectrum is often given
in terms of photon number emission spectrum, i.e. the number of photons
emitted at a particular frequency. In this case the expression in
Eq. \ref{Eq:PowSca} has to be divided by $\hbar\omega$ to convert
an energy to a photon number. 

Figure \ref{fig:Spectra_subplot} gives both electron energy loss
spectra and photon emission spectra for an electron passing the non-concentric
annulus vertically and horizontally. Both spectra are given as loss
probability density and photon emission probability density, i.e.
the area under the curves is unity. The total energy lost by the electron
and total energy scattered by the nano-particle is given in table
\ref{tb:ELossLeft_cresc}.

\begin{table}[t!]
\centering
\begin{tabular}{lllll}
  \toprule
  \midrule
    &\multicolumn{2}{c}{Vertical}  &\multicolumn{2}{c}{Horizontal}\\
    \cmidrule(r){2-3}        \cmidrule(r){4-5}
        & $e^-_{loss}$ $[eVm^{-1}]$ & $N(\gamma)$     &$e^-_{loss}$ $[eVm^{-1}]$    & $N(\gamma)$\\
  \midrule
  $c_e=0.05c$& $6.05\times10^{-11}$ & $3.76\times10^{-12}$ & $3.05\times10^{-11}$ & $8.71\times10^{-13}$\\
  $c_e=0.1c$ & $8.62\times10^{-11}$ & $5.69\times10^{-12}$ & $7.56\times10^{-11}$ & $3.63\times10^{-12}$\\
  $c_e=0.2c$ & $6.55\times10^{-11}$ & $4.15\times10^{-12}$ & $6.97\times10^{-11}$ & $3.65\times10^{-12}$\\
  \midrule
  \bottomrule
\end{tabular}
\caption{Table giving the total energy loss defined by $\int e_{loss}^{-}(\omega)d\omega$ in units of [eV per unit length] and the total number of photons emitted $N(\gamma)$. The different rows are for different electron velocities, the first two columns are for a (line) electron moving past the particle in the vertical direction and the last two columns are for a (line) electron moving horizontally past the particle. The geometrical parameters were set to $g^2=2\times10^{-8}$, $x_0=1.5$, $R_0=\exp(1)$ and $R_1=\exp(1.2)$, giving a particle of diamater $\approx 21nm$. The distance of the electron to the particle was $\approx3.3nm$.}\label{tb:ELossLeft_cresc}
\end{table}

In figure \ref{fig:Spectra_subplot} we consider an electron passing
the non-concentric annulus vertically (left) and horizontally (right).
We used experimental data for the permittivity of silver for the non-concentric
annulus \cite{JohnsonChristy}. The different curves correspond to
different electron velocities $c_{e}=0.05c,0.1c,0.2c$ with corresponding
kinetic energies of $\approx0.64eV,2.57eV,10.54eV$. 

The intuitive picture of what is happening is this. The electron moves
past the nano-particle very rapidly, transferring energy to the particle
over a very short window of time. The faster the electron, the smaller
this time window. Heisenberg's uncertainty principle $t_{\Delta}\omega_{\Delta}\approx\hbar$
then lets us estimate the spread of frequencies over which excitations
will happen (a few $eV$ in this case). The faster the electron, the
larger the range of frequencies present, meaning that we would expect
faster electrons to be better suited to excite higher frequency modes.
A conclusion which is supported by the source potential in Eq.\ref{Eq:PotSou},
which decays much faster for slower velocities or higher frequencies.
This is not what is observed. In the energy loss spectrum of the vertical
case the loss probability for the mode at $\omega\approx3.5$ is larger
for the electrons with $c_{e}=0.05$ and $c_{e}=0.1c$ than for the
fastest electrons. But this is not a universal behavior as the first
mode at $\omega\approx2.4eV$ is excited similarly for all three speeds,
whereas the mode at $\omega=2.5$ is entirely absent for the electron
with velocity $c_{e}=0.1c$. The results indicate that not all modes
of the system might be observable with electrons of arbitrary velocity. 

The reason for the vanishing of this mode is an `accidental degeneracy'.
While the formulae for the resistive losses and scattering spectrum
run over all modes of the system, they are all proportional to the
expansion coefficients of the source, $a_{n}^{s-}$. These are damped
oscillating functions with respect to the parameter $\omega/c_{e}$.
For the quadrupole mode at $\omega\approx2.5eV$ the zero of the expansion
coefficient $a_{2}^{s-}$ coincides with the resonance frequency of
this particular mode. Hence, the mode is not present in the exciting
source potential and it remains dark.

This accidental degeneracy does not have to be a nuisance, since it
allows the `switching off' of a particular mode by tuning the energy
of the exciting electrons. For example, switching off the mode at
$\omega\approx2.5eV$ for the electron with $c_{e}=0.1c$ resulted
in a much stronger excitation of the next order mode at $\omega\approx3.2$,
which is barely visible for the slower and faster electron. In the
horizontal case this is not as easily achieved, as the expansion coefficients
have both real and complex components, which means it is harder to
tune them to vanish at the same time in the relevant frequency range.
This explains why the spectra for the electrons moving horizontally
vary much more smoothly with respect to different electron velocities.
The other difference between the horizontal and vertical case is that
the higher order modes near the surface plasma frequency are excited
much more efficiently when the electrons move past the nano-particle
in the horizontal direction.

The photon emission spectra show similar behavior for the even modes
below the surface plasma frequency, but drop sharply (in both cases)
for the odd modes above the surface plasma frequency, as would be
expected.

\section*{Time-Response of non-concentric annulus}

The great advantage of the TO approach to EELS calculations is that
analytical expressions for the electrostatic potential $\phi(x',y',\omega)$
can be derived and easily computed. The time-evolution of these systems
can thus be efficiently reconstructed from the frequency domain via
a fourier transform
\begin{align}
\phi(x',y',t) & =\int_{-\infty}^{\infty}d\omega\phi(x',y'\omega)e^{-i\omega t}\\
 & =2\mbox{Re}\left[\int_{0}^{\infty}d\omega\phi(x',y'\omega)e^{-i\omega t}\right].
\end{align}
Here we used the reality condition on the potential $\phi(-\omega)=\phi^{*}(\omega)$
\cite{Jackson}. These calculations are fast enough to be carried
out on a state-of-the-art laptop within a couple of seconds (or minutes
depending on the number of spatial points), which makes it possible
to create videos of the time-evolution of these systems. See supplementary
files. For the time-evolution analysis we used a Drude model, $\epsilon_{m}=1-\omega_{p}^{2}/(\omega(\omega+i\gamma))$,
with $\gamma=0.32eV$ and $\omega_{p}=8$ for the permittivity of
the nano-particle and assumed the surrounding dielectric was air with
$\epsilon_{d}=1$.

An electron moving past a nano-particle excites many modes at once.
The number of excited modes and there relative strength can be inferred
from the energy electron loss spectrum. The different spectra for
electrons passing the particle in the vertical and horizontal direction
indicate that the time response for the vertical case will be dominated
by the first two (dipole and quadrupole) modes, while the one for
the horizontally moving electron is expected to be more complicated
with contributions from higher order modes as well. This is exactly
what can be observed in the videos provided as supplementary materials.
They show the time evolution of the vertical and horizontal electric
field components as the electron moves past the nano-particle. While
the electric field pattern observed for the vertical case only shows
one or two oscillations of the electric field as one moves around
the particle, the pattern for the horizontal case is much more complicated
and shows many nodes and anti-nodes, indicating that higher order
modes are contributing significantly to the scattered and induced
fields.
\begin{figure*}[t!]
\centering
\includegraphics[width=\textwidth]{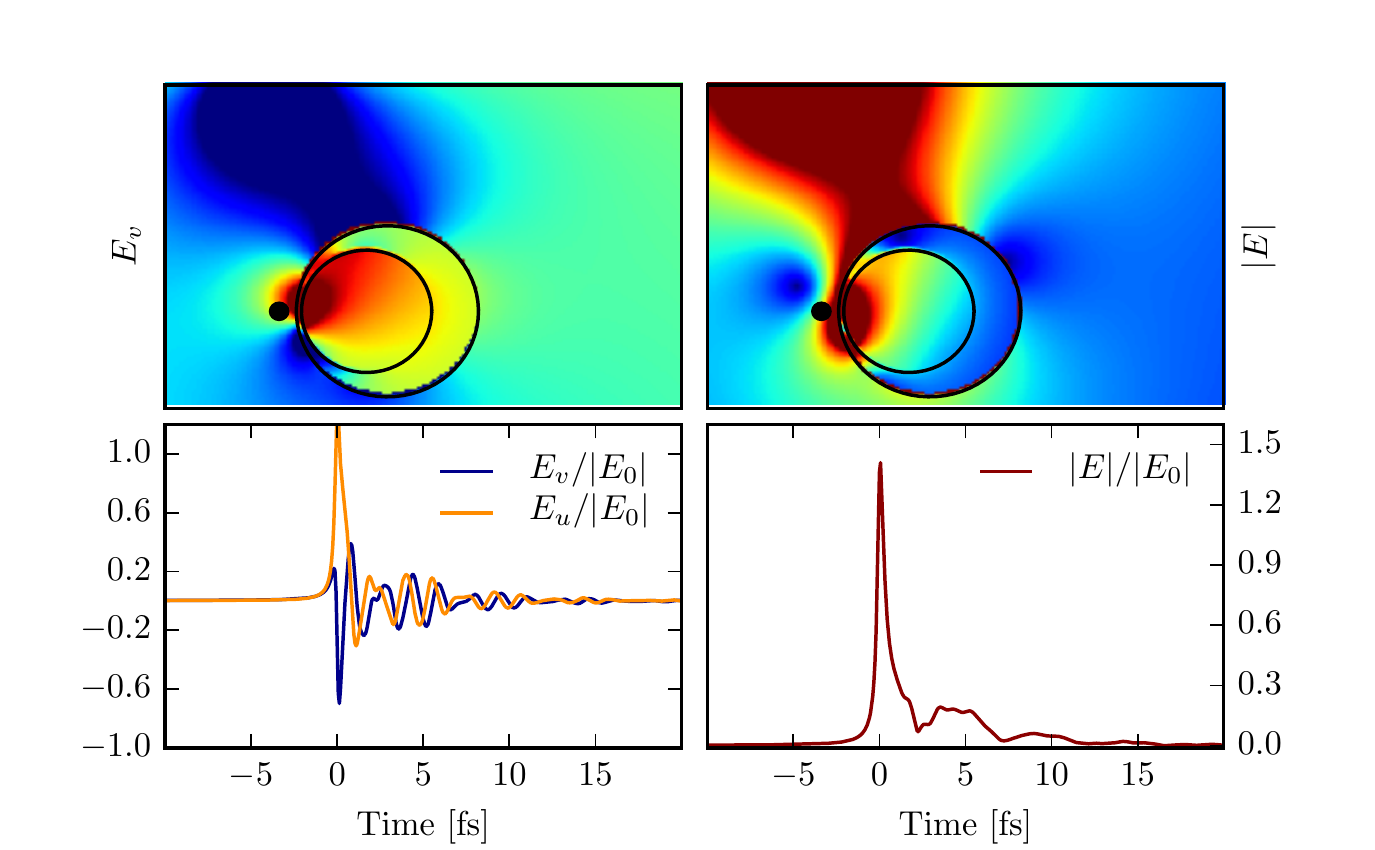}
\caption{The top panel shows the electric field distribution at a specific time point (indicated by the arrow). $E_v$ on the left and $|E|$ on the right. The bottom panel shows the time response of the crescent at a particular point (marked by the black dot in the top panel). In this case the electron moves vertically to the left of the nano-particle. The plots in the bottom panel have been normalised by the maximum source field at the particular point. The geometrical parameters were set to $g^2=2\times10^{-8}$, $x_0=1.5$, $R_0=\exp(1)$ and $R_1=\exp(1.2)$, giving a particle of diamater $\approx 21nm$. The distance of the electron to the particle was $\approx3.3nm$.}\label{fig:Time_plot_vert}
\end{figure*}

Singular nano-crescents have been shown to efficiently harvest light
and concentrate it at their singular point. This focusing of the energy
into a small area can also be observed here. For the electron passing
horizontally modes are excited at the `fat' end of the non-concentric
annulus, but can then be seen to propagate towards the thinnest point,
squeezing the energy into a tiny area. Unlike the singular crescent,
where the fields never reach the singularity, they can travel past
this point in the non-concentric annulus and travel on a circle until
all the energy is lost through scattering or resistive losses \cite{ZhangOEMultiFano, TOPlasmonics}.

\begin{figure*}[t!]
\centering
\includegraphics[width=\textwidth]{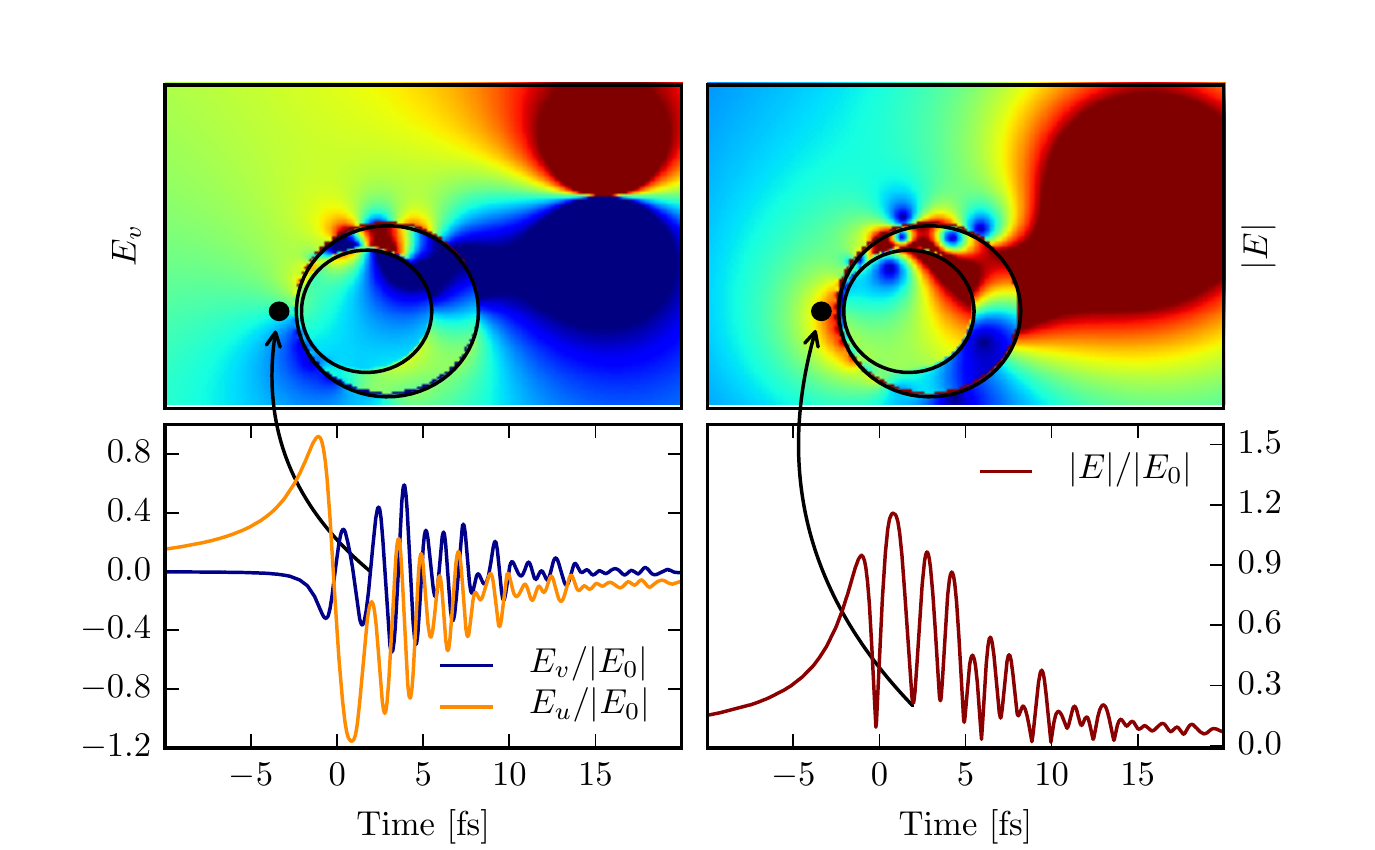}
\caption{The top panel shows the electric field distribution at a specific time point (indicated by the arrow). $E_v$ on the left and $|E|$ on the right. The bottom panel shows the time response of the crescent at a particular point (marked by the black dot in the top panel). In this case the electron moves horizontally along the top of the nano-particle. The plots in the bottom panel have been normalised by the maximum source field at the particular point. The geometrical parameters were set to $g^2=2\times10^{-8}$, $x_0=1.5$, $R_0=\exp(1)$ and $R_1=\exp(1.2)$, giving a particle of diamater $\approx 21nm$. The distance of the electron to the particle was $\approx3.3nm$.}\label{fig:Time_plot_hori}
\end{figure*}

Figure \ref{fig:Time_plot_vert} and \ref{fig:Time_plot_hori} give
snapshots of the video. The upper panels show contour plots of the
electric fields at a particular point in time, while the lower ones
give the time-evolution of the fields at a particular point in space.
The fields in the time-evolutions graphs have been normalized by the
maximum source field at that particular point. $t=0$ has been chosen
as the point in time when the electron moves through $x'=0$ (horizontal
case) and $y'=0$ (vertical case). As expected the time-evolution
looks more complicated for the horizontal case, with many more oscillations
in the fields than for the vertically passing electron. It should
also be noted that the response in the vertical case is much more
sudden, both in excitation and decay.

\pagebreak

\section*{Conclusion}

In this paper, we introduced a new approach based on Transformation
optics to calculate the frequency and time-domain response of plasmonic
systems under electron beam excitation. Transformation optics previously
proved to be a valuable tool in the analysis of the mode spectra and
optical response of plasmonic particle of complex geometries. Here,
we showed that it can also be employed to calculate the electron energy
loss and photon scattering spectra of plasmonic nano-particles under
electron beam excitation, with very good agreement between analytics
numerical simulations (see Supplementary Material). Moreover, due
to the analytical nature of this approach, it has been possible to
obtain the time-domain response of the nano-particle in a very time-efficient
manner by Fourier transforming the frequency domain solution. We believe
that Transformation optics is very well suited to analyze EELS and
CL systems in plasmonics and will give additional physical insight
into these.

\section*{Acknowledgements}
M.K. acknowledges support from the Imperial College PhD scholarship. 
Y.L. acknowledges NTU-A*start Silicon Technologies of Excellence under
the program grant No. 11235150003.
J.B.P. gratefully acknowledges support from the EPSRC EP/L024926/1
programme grant, the Leverhulme Trust and the Gordon and Betty Moore foundation.


\bibliographystyle{unsrt}
\bibliography{Main_bibliography}


\end{document}